\documentstyle[11pt]{article}

\input epsf
\begin{document}
\title{
Non-Markovian Dynamics and Entanglement in Quantum Brownian
Motion}
\author{
K. Shiokawa\thanks {E-mail address: kshiok@mail.ncku.edu.tw}
\\
{\small National Center for Theoretical Sciences,}\\
{\small National Cheng-Kung University, Tainan 701, Taiwan}}
\date{\today}
\maketitle
\begin{abstract}
Dynamical aspects of quantum Brownian motion in a low temperature
environment are investigated. We give a systematic calculation of
quantum entanglement among two Brownian oscillators without
invoking Born-Markov approximation widely used for the study of
open systems. Our approach is suitable to probe short time
dynamics at cold temperatures where many experiments on quantum
information processing are performed.\\


\end{abstract}
\newpage

\section{Introduction}

Chemical systems such as liquid NMR have been actively used for
quantum information processing
experiments\cite{LaflammerKnillNegrevergneMartinezSinhaCory02}.
Hybrid devices based on molecular ensembles on solid state circuits
provide many advantages as quantum processors such as long lasting
memory, good controllability and scalability\cite{ADDLMRSZ06}.

Implementing quantum information processing devices requires
precise knowledge of quantum open systems. The detailed study of
quantum open systems are often prevented by a limitation of
analytical or numerical resources to probe systems with large
degrees of freedom. The first principle calculation of open
quantum systems based on projection
operators\cite{Nakajima58,Zwanzig61} or influence
functionals\cite{FeyVer63} has been developed and applied to
numerous problems\cite{LCDFGZ87,Weiss99}. Reduced dynamics
obtained by these methods follows non-Markovian evolution carrying
the memory of environment. Solving the equations of motion
obtained by these method directly is generally difficult to deal
with as they are given by integro-differential equations.

Various approximation schemes have been used to simplify the exact
evolution to study open system evolution. Most commonly used
approximation is Born approximation in which the system-bath
interaction is treated perturbatively. However, naive Born
approximation applied to quantum open systems does not guarantee
the positivity of the density matrix and requires a caution
particularly when evaluating quantities sensitive to this aspect
such as quantum entanglement.

For a weak system-bath interaction, bath is disturbed by the
interaction only in a small amount and tends to reset to the
original state in a short time compared to the system time scale.
The correlation among the bath variables may become negligible
compared to the system time scale. In a Born-Markov approximation,
the bath correlation is ignored. At sufficiently low temperature,
however, the bath memory can persist for a long time and this
approximation is expected to break down. Regardless of the
temperature, the noise with long range correlation such as 1/f
noise cannot be treated under this approximation. 1/f noise is
ubiquitous causing a problem in implementing a quantum circuit
using electronic instruments in solid state, ion trap, or hybrid
molecular devices. Decoherence in a nonequilibrium
bath\cite{ShiokawaKapral02} also shows a peculiar behavior which
is significantly different from the one under Markov
approximation.

Further approximation such as rotating-wave approximation is
commonly used with Born-Markov approximation. There we assume weak
system-bath interaction such that a coupling becomes only among near
resonant modes. We further ignore counter rotating terms since they
are expected to be averaged out during a sufficiently long
observation time. The rotating-wave approximation is thus not
suitable for describing short time dynamics in a strong coupling
regime.

 Quantum entanglement,
manifestation of intrinsic nonlocality in quantum
mechanics\cite{EPR:35}, now became one of the most active research
topics in quantum information science. It also has been a long
historical issue as the discrepancy between quantum and classical
mechanics is a serious obstacle to understand macroscopic
classical and quantum mechanics from the view of microscopic
quantum dynamics\cite{WheelerZurek83,Dec96}.

Conversely but for the same reason, to manipulate quantum systems in
order to be useful in the macroscopic world is a formidable task.
Recent progress in this direction is motivated by rapid progress in
quantum technology, where there is a demand for a precise control of
small quantum devices. Quantum entanglement is considered to be a
valuable resource for practical applications such as quantum
computation and communication\cite{NielsenChuang00}.

Many works have been devoted to clarify the rigorous criteria for
entanglement. For continuous variables, the necessary and
sufficient criteria for entanglement can be given in terms of
Peres-Horodecki criteria\cite{Peres,Horodechi,Simon},
negativity\cite{VidalWerner02}, and entanglement of
formation\cite{GiedkeWolfKrugerWernerCirac03}. Most of these works
are devoted to study the static properties of general pure and
mixed states. In light of information processing, it is also
desirable to study the dynamical properties based on the realistic
models in an open system setting.

Quantized harmonic oscillators have been playing an important role
in a history of quantum mechanics. Realization of quantum
protocols based on continuous variables such as quantum
teleportation\cite{Furusawa98} and quantum key
distribution\cite{Grosshans03} show that they also play the
similar crucial role in quantum information science. Realization
of the similar protocols based on solid state devices is a highly
formidable task since the disturbance from environment in solid
state systems is much stronger than in optical systems.

So far the studies of continuous variables for open systems are
based on the master equation under Born-Markov rotating wave
approximation such as Lindblad equations\cite{RajagopalRendell01}
or the phenomenological model with high temperature Markovian
bath\cite{DoddHalliwell04}. Applications of these methods are
limited to the high temperature regime or slow dynamics with Ohmic
noise, while actual solid state implementation operates at low
temperatures manipulated with fast pulses and often suffers from
1/f noise.

Most of the solid state implementation of quantum information
devices operate in a ultra-cold temperature in order to maintain
quantum coherence. For instance, typical superconducting qubits
operate less than $100[mK]$\cite{DevoretMartinis04}. At ultra-cold
temperatures, the memory effect in the environment variables is no
longer negligible. The system-environment interaction in solid
state devices is typically much stronger than those qubits based
on atom-optical devices. These conditions are opposite to those
where the Born-Markov approximation is applicable. Thus we need to
develop the method to probe the regime beyond the conventional
Born-Markov approximation.

Born and rotating wave approximations are limited to weak coupling
regime and do not guarantee the positivity of the density matrix
evolution. Since separability criteria is based on the positivity
of the density matrix, a caution is required to study entanglement
under these approximations.

In the present paper, we study the precise mechanism of open
system entanglement dynamics based on a Brownian oscillator
model\cite{QBM0}, where both system and environment consist of
harmonic oscillators. Since our model is formally exactly
solvable, it makes us possible to probe the precise dynamics of
entanglement without conventional approximations. Previous works
for the study of a two-level system revealed that non-Markovian
dynamics that plays the major role at low temperatures is highly
nontrivial\cite{LCDFGZ87,ShiokawaHu04}.

In section 2, we develop systematic tools for our analysis. The
reduced density matrix in the Wigner representation is calculated
for two quantum Brownian oscillator model. In section 3, we first
make sure that uncertainty relations are always satisfied in our
model. They are time and temperature dependent as our system stays
mostly far from equilibrium. Next we make partial transpose
operation to our density matrix. In the Wigner representation, the
partial transpose operation corresponds to the partial mirror
reflection. We again study uncertainty relations after partial
transpose. According to the Peres-Horodecki-Simon criteria, the
violation of uncertainty relations after partial mirror reflection
can be used as a signature of quantum entanglement. Entanglement
measures, the negativity and the logarithmic negativity, are also
calculated for comparison.

\section{General formulation} \label{GF}

We consider a system composed of two harmonic oscillators. Our
Lagrangian is given by
\begin{eqnarray}
L_S = \sum_{j=1}^2 \frac{M_j \dot{R}_j^2}{2} - V(R_1,R_2)\;,
\label{HS}
\end{eqnarray}
where the potential $V$ is assumed to be harmonic:
\begin{eqnarray}
V(R_1,R_2)= \sum_{j=1}^2 \frac{M_j \Omega_{j}^2}{2} R_j^2.
\label{HS}
\end{eqnarray}
 Each oscillator variable
$R_j$ located at the origin $x=0$ couples linearly with a scalar
field $\phi$ via minimal coupling as
\begin{eqnarray}
L_I&=&-\sum_{j=1}^2 \lambda_{j}~ \dot{R}_{j}(t) \phi(0,t).
\label{HI}
\end{eqnarray}
The field Lagrangian is
\begin{eqnarray}
L_{\phi}&=&\frac{1}{2} \int d x \left[ (\partial_x \phi)^2 - m^2
\phi^2 \right].
\end{eqnarray}
The scalar field $\phi$ propagating in one-dimensional space allows
a mode decomposition:
\begin{eqnarray}
  \phi(x,t)=
  \int \frac{d k}{(2\pi)^{1/2}\sqrt{2 \omega_{k}}}
\left\{ b_k e^{-i \omega_k t + ikx} + b_k^{\dagger} e^{i \omega_k
t-ik x} \right\}.
 \label{field}
\end{eqnarray}
In this paper, we study the massless field, then $\omega_k = |k|$.
We consider the field $\phi$ as an environment and trace out to
obtain dissipative dynamics for oscillator variables $R_j$.

The Heisenberg equations that $R_j$ satisfy have the form of damped
 harmonic oscillators\cite{Weiss99}:
\begin{eqnarray}
M_j \frac{d^2 {R}_j(t)}{dt^2} + M_j \Omega_{j}^2 R_j(t) -
\sum_{l=1}^{2} \int_{0}^{t} ds \frac{d\alpha_{Ijl}(t,s)}{dt}
\frac{d R_l(s)}{ds} =0, \label{EL1}
\end{eqnarray}
where
\begin{eqnarray}
\alpha_{Ijl}(t,t')= -2 \lambda_{j} \lambda_{l} \sum_k \sin \left[
\omega_k (t-t')\right]/\omega_k \label{alphaI}
\end{eqnarray}
is an imaginary part of the response function\cite{FeyVer63} defined
as \\$\alpha_{jl}(t,t')\equiv 2\lambda_{j} \lambda_{l} \sum_k
e^{-i\omega_k (t-t')}/\omega_k$ with $\sum_k \equiv \int d
k/(2\pi)$.
 Note that $\alpha_{Ijl}(t,t')$ is antisymmetric in indices:
$\alpha_{Ijl}(t,t')=-\alpha_{Ilj}(t',t)$. Without any
approximations, Eq.(\ref{EL1}) has a nonlocal form with kernels
given by $\alpha_{Ijl}(t,t')$. Thus the value of $R_j$ at each
moment depends on their entire history of the past.

In one dimensional space when a ultraviolet cutoff of the field
modes is brought to infinity, Eq.(\ref{EL1}) will be reduced to a
local form. We write $\gamma_1\equiv
\lambda_1^2/M_1,\gamma_2\equiv \lambda_2^2/M_2,\gamma_{12} \equiv
\lambda_1 \lambda_2/M_{2},\gamma_{21} \equiv \lambda_1
\lambda_2/M_{1}$. In this case, Heisenberg equations of motion
will be
\begin{eqnarray}
M_1 \ddot{R}_1(t) &+& M_1 \Omega_{1}^2 R_1(t) + \gamma_1 M_1
\dot{R}_1(t)  + \gamma_{12}  M_2  \dot{R}_2(t)=0\nonumber\\ M_2
\ddot{R}_2(t) &+& M_2 \Omega_{2}^2 R_2(t) + \gamma_2 M_2
\dot{R}_2(t)  + \gamma_{21}  M_1 \dot{R}_1(t)=0.
 \label{ELNR}
\end{eqnarray}
We write a pair of solutions of (\ref{ELNR}) with initial conditions
$R_1(0)=R_2(0)=0$ and $\dot{R}_1(0)=1$, $\dot{R}_2(0)=0$ as $h_1(t)$
and $h_3(t)$. For identical two oscillators ($M_1=M_2=1$,
$\Omega_{1}=\Omega_{2}\equiv \Omega$) coupled to $\phi$ with equal
strength ($\lambda_1=\lambda_2$) (hereafter we will drop indices
from these parameters), the solutions are given by $h_1(t)\equiv
(g_{1}(t)+g_{0}(t))/2$ and $h_3(t)\equiv (g_{1}(t)-g_{0}(t))/2$,
where
\begin{eqnarray}
g_{1}(t) = \frac{\sin(\Omega_{r} t)}{\Omega_{r}}e^{-\gamma t}
~\mbox{and}~ g_{0}(t) = \frac{\sin(\Omega t)}{\Omega}
\label{solution2}
\end{eqnarray}
are solutions corresponding to two normal modes of a coupled
oscillator and $\Omega^2_r \equiv \Omega^2 - \gamma^2$.

General solutions with arbitrary initial conditions $R_{j0}$ and $P_{j0}$ of coupled
Heisenberg equations (\ref{ELNR}) for $j=1,2$ are
\begin{eqnarray}
R_j(t) &=&
   C_{R_{j}R_{1}} R_{10} +  C_{R_{j}P_{1}} P_{10}
  +C_{R_{j}R_{2}} R_{20} +  C_{R_{j}P_{2}} P_{20}
 \nonumber\\ &+&\lambda \int_{0}^{t} ds
g_{1}(t-s) \dot{\phi}(s),
\nonumber\\ P_j(t)  &=&
  C_{P_{j}R_{1}} R_{10} +  C_{P_{j}P_{1}} P_{10}
 +C_{P_{j}R_{2}} R_{20} +  C_{P_{j}P_{2}} P_{20}
\nonumber\\ &+&\lambda \int_{0}^{t} ds g_{2}(t-s) \dot{\phi}(s),
\label{ClassicalEvolution}
\end{eqnarray}
where ${g}_{2}\equiv\dot{g}_{1}$. The expectation value of phase
space variables can be expressed in a matrix form:
\begin{eqnarray}
\left( \begin{array}{c}
        \langle R_1 \rangle \\
        \langle P_1 \rangle \\
        \langle R_2 \rangle \\
        \langle P_2 \rangle
         \end{array}      \right)
         =
      {\cal C}
      \left( \begin{array}{c}
          R_{10} \\
          P_{10} \\
          R_{20} \\
          P_{20}
         \end{array}      \right)
=  \left( \begin{array}{cccc}
      C_{R_1 R_1}&C_{R_1 P_1}
     &C_{R_1 R_2}&C_{R_1 P_2}\\
       C_{P_1 R_1}&C_{P_1 P_1}
     &C_{P_1 R_2}&C_{P_1 P_2}\\
       C_{R_2 R_1}&C_{R_2 P_1}
     &C_{R_2 R_2}&C_{R_2 P_2}\\
       C_{P_2 R_1}&C_{P_2 P_1}
      &C_{P_2 R_2}&C_{P_2 P_2}\\
            \end{array}      \right)\left( \begin{array}{c}
          R_{10} \\
          P_{10} \\
          R_{20} \\
          P_{20}
         \end{array}      \right)
\label{RC2}
\end{eqnarray}
A time evolution matrix $ {\cal C}$ for our solutions in
(\ref{solution2}) is given by
\begin{eqnarray}
  {\cal C}
 \equiv
  \left( \begin{array}{cccc}
     f_1 & h_1 & f_3 & h_3\\
   f_2 & h_2 &
   f_4 & h_4 \\
   f_3 & h_3 & f_1 & h_1\\
   f_4 & h_4 & f_2 & h_2 \\
   \end{array}      \right),
     \label{C}
\end{eqnarray}
where $f_{2j-1}\equiv {h}_{2j}-\dot{h}_{2j}(0) g_{1}$ and
$f_{2j}\equiv \dot{f}_{2j-1}$ for $j=1,2$.

It is convenient for our purpose to introduce the Wigner
distribution function as
\begin{eqnarray}
  W(R_1,P_1,R_2,P_2) &=& \frac{1}{(2\pi)^2} \int d^2 r
  {\rho}(R_1-\frac{r_1}{2},R_2-\frac{r_2}{2},R_1+\frac{r_1}{2},R_2+\frac{r_2}{2})
  e^{i \sum_{j=1}^{2} P_j {r}_j}.\nonumber
   \label{Wigner}
    \end{eqnarray}
 The characteristic function\cite{MandelWolf95}
 for the Wigner distribution is given by
\begin{eqnarray}
  \chi_{W}({\cal Y},t)
  &=& \mbox{Tr} \left[ {\rho}(0)
  e^{i \sum_{j=1}^{4} Y_j {X}_j(t)}
            \right],
   \label{chiW}
    \end{eqnarray}
     where we
defined ${\cal X}\equiv(X_1 ... X_{4})$ with
   ${X}_{2j-1}\equiv \sqrt{\Omega} {R}_j$,
   ${X}_{2j}\equiv  {P}_j/\sqrt{\Omega}$
   for $j=1,2$
   and ${\cal Y}\equiv(Y_1 ... Y_{4})$.
   We will fix the normalization for each component that
   appeared in $ {\cal C}$ accordingly.
  The symmetric correlations can be obtained from $\chi_{W}({\cal Y},t)$ as
   \begin{eqnarray}
   \langle \{ X_i, X_j \} \rangle
  &=& -\frac{\partial^2 \chi_{W}({\cal Y},t)} {\partial {Y_i}\partial {Y_j}}|_{{\cal Y}=0},
   \label{XiXj}
    \end{eqnarray}
 where   $ \{A,B\} \equiv (AB+BA) /2$ is an anticommutator.
We trace out the field $\phi$ in order to obtain the reduced
dynamics of the system. With a factorized initial condition:
${\rho}(0) =
  {\rho}_{S}(0) \otimes {\rho}_{\phi}(0)$, $\chi_{W}$ is also factorized
to two components as
  $ \chi_{W}({\cal Y},t) = \chi_{W}^S({\cal Y},t) \chi_{W}^{\phi}({\cal Y},t)$.
In our case, the system part is $\chi_{W}^S({\cal
Y},t)=\mbox{Tr}_{S} \left[ {\rho}_S(0) e^{i \sum_{j=1}^{4} Y_j
{X}_{Cj}(t)}
            \right] $, where ${X}_{Cj}(t)$ are solutions of Heisenberg
            equations with $\phi=0$.
The field characteristic function in (\ref{chiW}) can also be
evaluated exactly. We assume that environment is initially in a
thermal state with an inverse temperature $\beta \equiv 1/T$. Its
density matrix is given as ${\rho}_{\phi}(0)=\sum_{k} e^{-\beta
\omega_k} \mid k\rangle \langle k\mid$. We obtain
\begin{eqnarray}
\chi_{W}^{\phi}({\cal Y},t) &=& \exp\left[ -\frac{1}{2} {\cal Y}^T
      {\bf \Large \Sigma} {\cal Y}
\right]\nonumber\\ &=& \exp\left[ -\frac{1}{2}(Y_1 ... Y_{4})^T
  \left( \begin{array}{ccc}
      \Sigma_{11} & ... & \Sigma_{14}   \\
       & ... &  \\
          \Sigma_{41} & ... & \Sigma_{44}
            \end{array}      \right)
        \left(   \begin{array}{c}
        Y_1 \\ ...\\ Y_{4} \\
         \end{array}      \right)
\right].
   \label{chiBGauss}
\end{eqnarray}
Here
\begin{eqnarray}
\Sigma_{jl}(t) &=& \frac{\lambda^2}
  { 2 \pi } \int_{0}^{\infty} d\omega
  \omega e^{-\omega/ \Lambda} \coth(\beta \omega/2) \int_{0}^{t} ds \int_{0}^{t}
ds' g_{\tilde{j}}(t-s) \cos \omega (s-s') g_{\tilde{l}}(t-s'),
\nonumber \\
   \label{sigma}
\end{eqnarray}
where $\tilde{j}\equiv(3+(-1)^j)/2$, are time-dependent
(nonequilibrium) fluctuations of the system variables, a part
induced from environment. Here we introduced the cutoff frequency
$\Lambda$ for the field modes.
 Note that off-diagonal correlations
$\Sigma_{jl}$ for $j\neq l$ are nonvanishing in general, that is,
an interaction with a common environment induces an effective
interaction between two oscillator variables and thus correlation
and entanglement between them.

For an initial Gaussian state with vanishing mean positions and
momentums, $\langle{\cal X}(0)\rangle=0$, the system characteristic
function also takes a Gaussian form:
\begin{eqnarray}
\chi_{W}^S({\cal Y},t)&=& \exp\left[ -\frac{1}{2}{\cal Y}^T
   (\Delta {\cal X})_{C}^2(t) {\cal Y}
  \right] \\
  &\equiv& \exp\left[ -\frac{1}{2}(Y_1 ... Y_{4})^T
  \left( \begin{array}{ccc}
\langle  \{X_{1C}, X_{1C}\}  \rangle &...&\langle \{
X_{C1}, X_{C 4}\}\rangle \\& ... &\\
\langle \{X_{C 4}, X_{C1}\}\rangle
   &...&\langle  \{X_{C 4}, X_{C 4}\}\rangle
    \end{array}\right)
       \left(   \begin{array}{c}
        Y_1 \\ ...\\ Y_{4} \\
         \end{array}      \right)
\right], \nonumber
   \label{chiBGauss}
\end{eqnarray}
where ${\cal X}_{C}=(X_{C1} ...X_{C4})$ satisfy the equations of
motion (\ref{ELNR}) for damped harmonic oscillators. $(\Delta {\cal
X})_{C}^2(t)$ are essentially the initial fluctuations of system
variables shifted by damped oscillatory motion of a coupled harmonic
oscillator. Combining with the characteristic function for the
field, we obtain
\begin{eqnarray}
 \nonumber\\
& &W({\cal X},t) = \frac{1}{(2\pi)^2}\frac{1}{(\mbox{det} (\Delta
{\cal X})^2(t) )^{1/2}}\exp\left[ -\frac{1}{2}{\cal X}^T
 ((\Delta {\cal X})^2(t))^{-1}
 {\cal X}
  \right],  \nonumber
   \label{WignerFinal}
\end{eqnarray}
where $(\Delta {\cal X})^2=(\Delta {\cal X})_{C}^2+\Sigma$.

\section{Entanglement dynamics of quantum Brownian oscillators}

Let us consider a two-mode squeezed state with a squeezing parameter
$r$ as an initial state\cite{QOtext}. Its correlation matrix is
\begin{eqnarray}
(\Delta {\cal X})_{C}^2(0)\equiv \langle  \{{\cal X}_{C}(0),{\cal
X}^{T}_{C}(0) \}  \rangle &=&
\frac{1}{2}  \left( \begin{array}{cc} \cosh(2r) {\bf  1} & -\sinh(2r)\sigma_3 \\
-\sinh(2r) \sigma_3 & \cosh(2r) {\bf 1}
                \end{array}      \right).\nonumber
               \label{Cmatrix}
\end{eqnarray}
In the Wigner representation, the same state can be expressed as
\begin{eqnarray}
W(R_1,R_2,P_1,P_2)=\frac{4}{\pi^2} e^{-e^{2r}\left[ \Omega
(R_1-R_2)^2+(P_1+P_2)^2/\Omega\right] -e^{-2r}\left[
\Omega(R_1+R_2)^2+(P_1-P_2)^2/\Omega\right]}.
               \label{WEPR}
\end{eqnarray}  This state can be
obtained by acting a squeezing operator $e^{ir(R_1P_2-P_1R_2) }$ on
the vacuum.

A criteria for separability of a bipartite two-level-system was
studied in \cite{Peres}. The necessary and sufficient condition for
separability of the density matrix is to have only non-negative
eigenvalues after partial transpose of one of its subsystem. The
same criteria is not always sufficient for the bipartite system with
more than two levels\cite{Horodechi}. The extension of this criteria
to continuous Gaussian variables was first given in \cite{Simon}.
For Gaussian variables, the partial transpose of a density matrix in
a coordinate representation for one oscillator component is
equivalent to a mirror reflection of that component in the Wigner
distribution function. The necessary and sufficient condition for a
continuous variable quantum state to be separable is that the
partial mirror reflected state is still a physical quantum state
that satisfies the uncertainty principle. In the phase space
representation, the partial mirror reflection on the second variable
can be stated as $(R_1,P_1,R_2,P_2)\rightarrow(R_1,P_1,R_2,-P_2)$.
In terms of ${\cal X}$, this can be expressed as the matrix
operation by the matrix $\eta\equiv \mbox{diag}(1,1,1,-1)$ as ${\cal
X}\rightarrow \eta {\cal X}$. It follows that the partial mirror
reflection transforms the covariance matrix as
\begin{eqnarray}
(\Delta {\cal X})^2 \rightarrow \eta (\Delta {\cal X})^2
\eta^T.\label{XbyPT}\end{eqnarray}

Before we apply the above criteria to our covariance matrix
$(\Delta {\cal X})^2 $, let us make three local Bogoliubov
transformations to simplify the form of the covariance matrix.
These local transformations do not change the separability of the
system. First we consider the local orthogonal transformation
\begin{eqnarray} M_4 \equiv \left(
\begin{array}{cc}
   M_2   & 0    \\
   0     & M_2  \end{array} \right) \in O(2,R)\bigotimes
O(2,R)\subset O(4,R),
          \label{Olocal}
\end{eqnarray}
where $M_2 \in O(2,R)$ is an orthogonal matrix. Under $M_4$, a
symmetric matrix of the form
\begin{eqnarray}
(\Delta {\cal X})^2 \equiv \langle \{\cal{X},\cal{X}^{T}\} \rangle
\equiv \left(
\begin{array}{cc}
   D   & A    \\
   A^{T} & D  \end{array}      \right)
    \label{V}
\end{eqnarray}
transforms to
\begin{eqnarray} M_4 (\Delta {\cal X})^2 M_4^T \equiv \left(
\begin{array}{cc}
   M_2 D  M_2^T   &  M_2 A  M_2^T   \\
   M_2 A^{T}  M_2^T & M_2 D M_2^T  \end{array}      \right).
         \label{SVS}
\end{eqnarray}
By  a suitable choice of $M_4$, we can diagonalize $D$. Next we
make a local symplectic transformation with another matrix $S_4
\in Sp(2,R)\bigotimes Sp(2,R) $ that has a form:
\begin{eqnarray} S_4 \equiv \left(
\begin{array}{cc}
   S_2   & 0    \\
   0     & S_2  \end{array}      \right),
           \label{Slocal}
\end{eqnarray}
where $S_2 \in Sp(2,R)$ is a symplectic matrix. For a suitable
choice of $S_2$, we can make $ S_2~ M_2 D  M_2^T ~S_2^T $ to be
diagonal with an equal component $d$.
 Furthermore another transformation with an orthogonal
matrix $ O_4 \in O(2,R)\bigotimes O(2,R) $ that has a form:
\begin{eqnarray} O_4 \equiv \left(
\begin{array}{cc}
   O_1   & 0    \\
   0     & O_2  \end{array}      \right)
           \label{matO}
\end{eqnarray}
can make $(\Delta {\cal X})^2$ into the following canonical form:
\begin{eqnarray} (\Delta {\cal X}_C)^2 = \left(
\begin{array}{cccc}
   d & 0 & a & 0  \\
   0 & d & 0 & b  \\
   a & 0 & d & 0  \\
   0 & b & 0 & d  \\
\end{array}      \right).
           \label{canonicalV}
\end{eqnarray}

From the Williamson's theorem\cite{Williamson36}, there exists a
symplectic transformation that diagonalizes a positive-definite $4
\times 4$ symmetric matrix into the following form:
\begin{eqnarray}
  (\Delta {\cal X}_D)^2 = \left( \begin{array}{cccc}
   \zeta_1 & 0       &0    & 0  \\
      0    & \zeta_1 & 0 & 0 \\
      0    & 0& \zeta_2 & 0     \\
      0    & 0& 0     & \zeta_2     \end{array}      \right).
      \nonumber
   \label{DXdiag}
\end{eqnarray}
Although such a symplectic transformation does not preserve the
eigenvalue spectrum in general, the diagonal components $\zeta_l$
for $l=1, 2$ can be calculated. Writing a commutation relation in a
matrix form as $\left[X_i,X_j\right]=i\Gamma_{ij}$ with
\begin{eqnarray}
 \Gamma = \left( \begin{array}{cccc}
  0 & 1 & 0 & 0  \\
 -1 & 0 & 0 & 0 \\
  0 & 0 & 0 & 1     \\
  0 & 0 & -1 & 0 \end{array}      \right),\nonumber
   \label{Gamma}
\end{eqnarray}
we construct a real symmetric matrix $\Delta {\cal X} \Gamma
(\Delta {\cal X})^2 \Gamma^{T} \Delta {\cal X}$. This matrix has
an eigenvalue spectrum $\zeta_l^2$ ($l=1,2$)\cite{SMD:94}. The
uncertainty relation can be generalized to a symplectic invariant
form $(\Delta {\cal X})^2+i\Gamma/2\geq0$. By changing to the
diagonalized form $(\Delta {\cal X}_D)^2$ above, the uncertain
relation is equivalent to saying that $\zeta_l \geq 1/2$ for all
$l$.  Although this uncertain relation is invariant in arbitrary
symplectic transformations, they can change the entanglement
property. Thus we restrict our transformation to local symplectic
transformations and use the canonical form of $(\Delta {\cal
X}_C)^2$ in (\ref{canonicalV}) for our analysis. Our separability
criteria is invariant under local transformations.

The uncertainty relation expressed by the components in $(\Delta
{\cal X}_C)^2$ are given by
\begin{eqnarray}
\begin{array}{c}
(d+a)(d+b)  \geq \frac{1}{4} \\
(d-a)(d-b)  \geq \frac{1}{4} \end{array}.
           \label{Slocal}
  \label{UCR}
\end{eqnarray}
These are generalizations of the familiar uncertain relations for
pure state two oscillators to general mixed states. They can be
expressed as $\Delta \tilde{R}_1^2 \Delta \tilde{P}_1^2 \geq
\frac{1}{4}$ and
 $\Delta \tilde{R}_2^2 \Delta \tilde{P}_2^2 \geq \frac{1}{4}$
in the coordinates that diagonalize the correlation matrix
$(\Delta {\cal X})^2$.
\begin{figure}[h]
 \begin{center}
\epsfxsize=0.8\textwidth \epsfbox{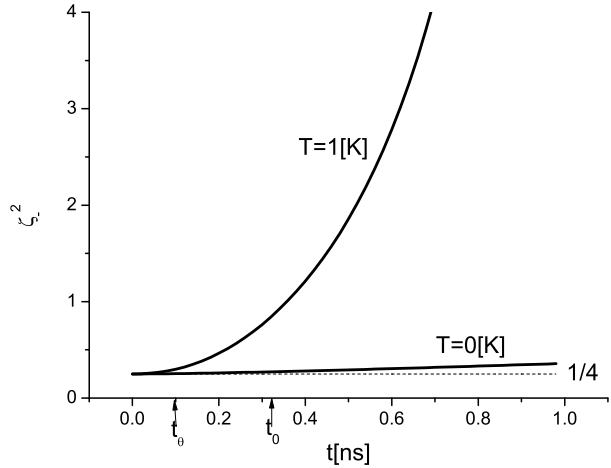}
\end{center}
\caption{ The temporal evolution of the uncertainty function
 ( $\zeta_{-}$ in Eqn. (\ref{UCRinv}) ) is plotted.
 The initial condition is a two mode squeezed state with $r=0.05$.  Other
 parameters are
$\gamma=0.01[GHz]$,~$\Lambda=50[GHz]$,~$\Omega=1.0[GHz]$.
\label{fig1}}
\end{figure}

 Under the partial transpose in (\ref{XbyPT}), $b \rightarrow -b$.
Thus the conditions for separability can be written as
\begin{eqnarray}
\begin{array}{c}
(d+a)(d-b)  \geq \frac{1}{4} \\
(d-a)(d+b)  \geq \frac{1}{4}. \end{array}
  \label{SEP}
\end{eqnarray}
One can easily see that the two-mode squeezed state we introduced in
(\ref{WEPR}) satisfies the uncertainty relation in Eq.(\ref{UCR}).
The separability condition (\ref{SEP}) implies $e^{4r} \geq 1$ and $
e^{-4r} \geq 1$, which only holds if $r=0$. Thus this state is
entangled as long as $r\neq 0$.
 In $r \rightarrow \infty$ limit, the state
becomes the original EPR state discussed in \cite{EPR:35}.

In order to make invariance under the local Bogoliubov
transformation manifest, one can write uncertainty relations
(\ref{UCR}) explicitly by using the symplectic invariants
constructed from the determinants of covariances $|A|$, $|D|$,
$|(\Delta {\cal X})^2|$ as
\begin{eqnarray}
\begin{array}{c}
\zeta_{\pm}^2 = |D|+|A|\pm \sqrt{(|D|+|A|)^2-|(\Delta {\cal
X})^2|}
 \geq \frac{1}{4}.
\end{array}
  \label{UCRinv}
\end{eqnarray}
In Fig. 1, the temporal behavior of the uncertainty function
$\zeta_{-}$ is plotted. The initial state is a pure two mode
squeezed state and satisfies the minimum uncertainty $1/4$. As the
state becomes mixed, the uncertainty increases monotonically in
time even for a zero temperature case. At higher temperature, the
rate of increase is faster.

Similarly the separability conditions (\ref{SEP}) are
\begin{eqnarray}
\begin{array}{c}
\lambda_{\pm}^2 = |D|-|A|\pm \sqrt{(|D|-|A|)^2-|(\Delta {\cal
X})^2|}
 \geq \frac{1}{4}.
\end{array}
  \label{SEPinv}
  \end{eqnarray}
Note that the inequalities for $\zeta_{+}$ and $\lambda_{+}$ in
(\ref{UCRinv}) and (\ref{SEPinv}) follow automatically from those
for $\zeta_{-}$ and $\lambda_{-}$. Thus $\lambda_{-}$ carries the
essential information on the separability of quantum states. In
Fig. 2, time evolution of the $\lambda_{-}$ is plotted. For an
initial coherent state, the uncertainty relation for the partial
transposed state is always satisfied throughout the whole
evolution indicating that there is no entanglement. For an initial
squeezed state ($r=0.1$), the uncertainty relation is violated
initially but eventually satisfied indicating that there is a
crossover from an entangled to a separable state. The asymptotic
value of separability seen in $\lambda_{-}$ appears to be
independent of the degree of initial squeezing.
\begin{figure}[h]
 \begin{center}
\epsfxsize=0.8\textwidth \epsfbox{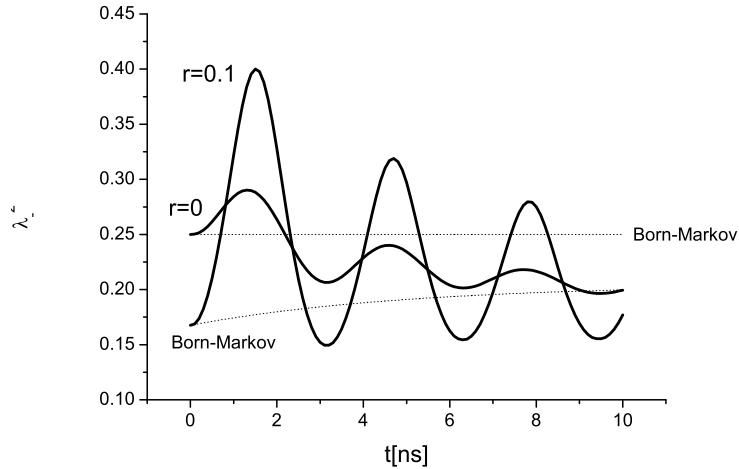}
\end{center}
\caption{ The temporal evolution of the uncertainty function
 ( $\lambda_{-}$ in Eqn. (\ref{SEPinv}) ) after partial transpose is plotted.
 The initial condition is a two mode squeezed state with $r=0.1$ for the thick solid curve
 and a coherent state ($r=0$) for the thin solid curve.  Other
 parameters are
$\gamma=0.1[GHz]$,~$\Lambda=50[GHz]$,~$\Omega=1.0[GHz]$,~$T=0$.
\label{fig2}}
\end{figure}

The negativity $\cal{N}$\cite{VidalWerner02,ASI04} for quantum
Brownian particles can be defined as
\begin{eqnarray}
\cal{N}&=&\frac{||\rho_r^{T}||-1}{2},
  \label{negativity}
   \end{eqnarray}
where $\rho_r^{T}$ is the reduced density matrix after partial
transpose. $\cal{N}$ is equal to the sum of all negative eigenvalues
of $\rho_r^{T}$ and measures how much $\rho_r^{T}$ fails to be
positive. From the Peres criteria, it can be used as a measure of
entanglement. It also has a nice property as an entanglement
monotone such that it does not increase under local operations and
classical communications. The logarithmic negativity $E_{\cal{N}}$
defined as
\begin{eqnarray}
E_{\cal{N}}&=&\log_2 ||\rho_r^{T}||
  \label{lognegativity}
   \end{eqnarray}
   also has the similar property.
Since diagonalization of $(\Delta {\cal X})^2$ brings the
   original state into the thermal state, the partial
   transposed density matrix after the same transformation
   also has the thermal form that can be written as a function of
   the symplectic invariants $\lambda_{\pm}$ as
   \begin{eqnarray}
\rho_r^{T} &=& \prod_{\pm} \left[ \left(\frac{2}{2\lambda_{\pm}+1}
\right)
 \sum_{n=1}^{\infty} \left(\frac{2\lambda_{\pm}-1}{2\lambda_{\pm}+1}
 \right)^n
|n_{\pm} \rangle \langle n_{\pm} | \right].
  \label{rhoTthermal}
   \end{eqnarray}
   For separable states, $\lambda_{\pm}\geq
   1/2$. Thus $||\rho_r^{T}||=1$ and ${\cal N}=E_{\cal N}=0$ follows.
   For entangled states, $\lambda_{-}<1/2$ but $\lambda_{+}\geq
   1/2$. The latter follows because $\lambda_{+}>
   \zeta_{-}$ for $|A|<0$ (if $|A|>0$, the state is separable
   from (\ref{Slocal}) to (\ref{SEPinv})).
   $||\rho_r^{T}||=1/2\lambda_{-}$ follows.
   Thus both $\cal{N}$ and
$E_{\cal{N}}$ can be expressed in terms of $\lambda_{-}$ as
\begin{eqnarray}
\cal{N}&=&\mbox{max}
\left[0,\frac{1-2\lambda_{-}}{4\lambda_{-}}\right], \nonumber\\
E_{\cal{N}}&=&\mbox{max}\left[0, -\log (2 \lambda_{-})\right].
  \label{negativity}
   \end{eqnarray}
\begin{figure}[h]
 \begin{center}
\epsfxsize=0.8\textwidth \epsfbox{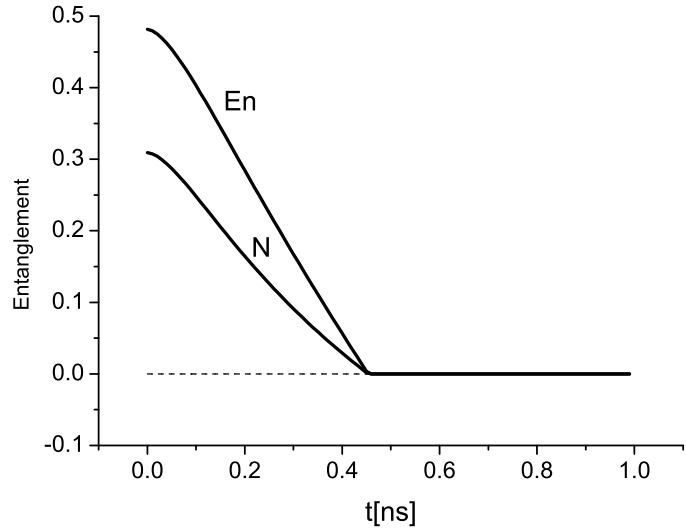}
\end{center}
\caption{ The negativity ${\cal{N}}$ and the logarithmic negativity
$E_{\cal{N}}$ are shown. The initial state is a two-mode squeezed
state with $r=0.1$. ${\cal{N}}$ and $E_{\cal{N}}$ both remain zero
in case of the two mode coherent state initial condition $r=0$.
Other parameters are
$\gamma=0.1[GHz]$,~$\Lambda=50[GHz]$,~$\Omega=1.0[GHz]$,~$T=0$.
\label{fig3}}
\end{figure}
In Fig. 3, the negativity ${\cal{N}}$ and the logarithmic
negativity $E_{\cal{N}}$ are shown as a function of time. The
initial pure two mode squeezed state has the maximum entanglement
that decays monotonically in time. They both vanish at
$t_{DE}=0.45[ns]$ and remain zero. This disentanglement time scale
$t_{DE}$ measured this way is the same as the time when the
uncertainty relation for the partial transposed state recovers
(the lower curve in Fig. 2) as we expect.

\section{Summary}

In this work, we used two quantum Brownian oscillator model to
study the dynamical aspect of quantum entanglement without
Born-Markov approximation. We studied several different criteria
for quantum entanglement. The uncertainty function for the
two-mode squeezed state under partial transpose initially violates
the uncertainty principle but eventually satisfies it. Invoking
the Peres-Horodecki-Simon's criteria, this corresponds to the
temporal crossover from an entangled to separable state. The
negativity and the logarithmic negativity show a monotonic
decrease and vanish indicating the similar crossover. We thus saw
that, through the analysis of exact dynamics, the effect of
environment destroys quantum entanglement among Brownian
oscillators through the decoherence mechanism.


\section*{Acknowledgments}
This work is supported by National Research Council, Taiwan: NSC
96-2119-M-007-001.


\end{document}